\documentclass[useAMS,usenatbib,usegraphicx]{mn2e}

\usepackage{epsf,amsmath,rotating}

\def\Teff{{T_{\rm eff}}}
\def\VT{{V_{\rm T}}}
\def\apj{ApJ}
\def\apjs{ApJS}
\def\aj{AJ}
\def\mnras{MNRAS}
\def\aa{A\&A}

\title[Cool White Dwarfs in SDSS Stripe 82]{\sc New UltraCool and Halo
  White Dwarf Candidates in SDSS Stripe 82}

\author[S. Vidrih et al.]{S. Vidrih$^{1,2}$,
        D.\ M. Bramich$^{1,3}$,
		  P.\ C.\ Hewett$^1$,
		  N.\ W.\ Evans$^1$,
		  G.\ Gilmore$^1$,
        \newauthor
        S.\ Hodgkin$^1$,
		  M.\ Smith$^1$,
		  L.\ Wyrzykowski$^1$,
        V.\ Belokurov$^1$,
        M.\ Fellhauer$^1$,
        \newauthor
		  M.\ J.\ Irwin$^1$,
        R.\ G.\ McMahon$^1$,
        D.\ Zucker$^1$,
        J.\ A.\ Munn$^4$, 
		  H.\ Lin$^5$,
		  \newauthor
		  G.\ Miknaitis$^5$,
		  H.\ C.\ Harris$^4$,
		  R.\ H.\ Lupton$^6$,
		  D.\ P.\ Schneider$^7$.\\
  $^{1}$ Institute of Astronomy, University of Cambridge, Madingley
  Road, Cambridge CB3 0HA, UK \\
  $^{2}$ Astronomisches Rechen-Institut/Zentrum f\"ur Astronomie der 
  Universit\"at  Heidelberg, M\"onchhofstrasse 12-14, 69120 Heidelberg, \\
  Germany \\
  $^{3}$ Isaac Newton Group of Telescopes, Apartado de Correos 321,
       E-38700 Santa Cruz de la Palma, Canary Islands, Spain \\
  $^{4}$ US Naval Observatory, 10391 West Naval Observatory Road, Flagstaff, AZ
  86001-8521, USA\\
  $^{5}$ Fermi National Accelerator Laboratory, P.O. Box 500, Batavia,
  IL 60510, USA \\
  $^{6}$ Princeton University Observatory, Princeton, NJ 08544, USA\\
  $^{7}$ Department of Astronomy and Astrophysics, Pennsylvania
  State University, 525 Davey Laboratory, University Park, PA 16802, \\
  USA}

\begin{document} 

\maketitle

\begin{abstract}

A $2.5^\circ \times 100^\circ$ region along the celestial equator (Stripe 82)
has been imaged repeatedly from 1998 to 2005 by the Sloan Digital Sky Survey. A
new catalogue of $\sim 4$ million light-motion curves, together with over 200
derived statistical quantities, for objects in Stripe 82 brighter than $r\sim
21.5$ has been constructed by combining these data by Bramich et al. (2007). 
This catalogue is at present the deepest catalogue of its kind. Extracting the 
$\sim 130\,000$ objects with highest signal-to-noise ratio proper motions, we 
build a reduced proper motion diagram to illustrate the scientific promise of 
the catalogue. In this diagram disk and halo subdwarfs are well-separated from 
the cool white dwarf sequence. Our sample of 1049 cool white dwarf candidates
includes at least 8 and possibly 21 new ultracool H-rich white dwarfs 
($\Teff < 4000$\,K) and one new ultracool He-rich white dwarf candidate 
identified from their SDSS optical and UKIDSS infrared photometry. At least 10 
new halo white dwarfs are also identified from their kinematics.

\end{abstract}


\section{Introduction}

The number of detected cool white dwarfs has risen dramatically with the advent
of the deep all-sky surveys, like the Sloan Digital Sky Survey (SDSS). 
Nonetheless, certain classes of white dwarfs -- for example, ultracool white 
dwarfs and halo white dwarfs -- remain intrinsically scarce.

The first ultracool ($\Teff < 4000$\,K) white dwarfs were discovered 
by~\citet{Ha99} and \citet{Ho00}.  Molecular hydrogen in their atmospheres 
causes a high opacity at infrared wavelengths, producing a spectral energy 
distribution with depleted infrared flux~\citep[e.g.,][]{Ha99,Be02}. Six 
ultracool white dwarfs have already been found in the SDSS on the basis of their
unusual colours and spectral shape~\citep{Ha01,Ga04}. Recently, \citet{Ki06} 
used a combination of SDSS photometry and United States Naval Observatory `B'
(USNO-B) astrometry to double the still tiny sample of these interesting 
objects, which probe the earliest star formation in the Galactic disk.

The halo white dwarfs are of astronomical interest as probes of the earliest
star formation in the proto-Galaxy, and as tests of the age of the oldest stars.
\citet{Li98} first identified six candidate halo white dwarfs on the basis of 
high proper motions. Subsequently, \citet{Op01} claimed the discovery of 38 high
proper motion white dwarfs, but it is unclear whether they are halo or thick 
disk members~\citep{Re01}. \citet{Ha06} recently identified a sample of 
$\sim 6000$ cool white dwarfs from SDSS Data Release 3 (DR3) using reduced 
proper motions, based on SDSS and USNO-B combined data~\citep{Mu04}. The sample
included 33 objects with substantial tangential velocity components ($>$ 160 
kms$^{-1}$) and so are excellent halo white dwarf candidates.

In this paper, we also use SDSS data to identify new members of the cool white
dwarf population. SDSS Stripe 82 is a $2.5^\circ \times 100^\circ$ strip along
the celestial equator which has been repeatedly imaged between 1998 and 2005.
We exploit the new catalogue of almost four million light-motion curves in 
Stripe 82~\citep{Br07}. By extracting the subset of $\sim 130\,000$ objects with
high signal-to-noise ratio proper motions, we build a clean reduced proper
motion diagram, from which we identify 1049 cool white dwarfs up to a magnitude 
$r\sim 21.5$\footnote{Magnitudes on the AB-system are used throughout this 
paper.}. Further diagnostic information is obtained by combining the catalogue 
with near-infrared photometry from the UKIRT Infrared Digital Sky Survey 
(UKIDSS) Data Release 2 (DR2) \citep{Wa07b}. This enables us to present new, 
faint samples of the astronomically important ultracool white dwarfs and halo 
white dwarfs.

\begin{figure}
\includegraphics[width=\hsize]{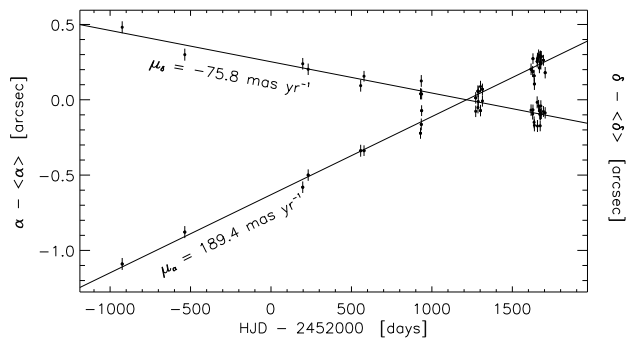}
\caption{Measured proper motion in RA (track at bottom left, left-hand y-scale) 
	and Dec (track at top left, right-hand y-scale) for the ultracool white dwarf
	candidate SDSS J224845.93--005407.0.
	\label{fig:pm}}
\end{figure}

\section{SDSS and UKIDSS Data on Stripe 82}

SDSS is an imaging and spectroscopic survey~\citep{Yo00} that has mapped more 
than a quarter of the sky. Imaging data are produced simultaneously in five 
photometric bands, namely $u$, $g$, $r$, $i$, and 
$z$~\citep{Fu96,Gu98,Ho01,Am06,Gu06,Am07}. The data are processed through 
pipelines to measure photometric and astrometric 
properties~\citep{Lu99,St02,Sm02,Pi03,Iv04,Tu06}.

\begin{figure*}
\begin{center}
\includegraphics[width=\hsize]{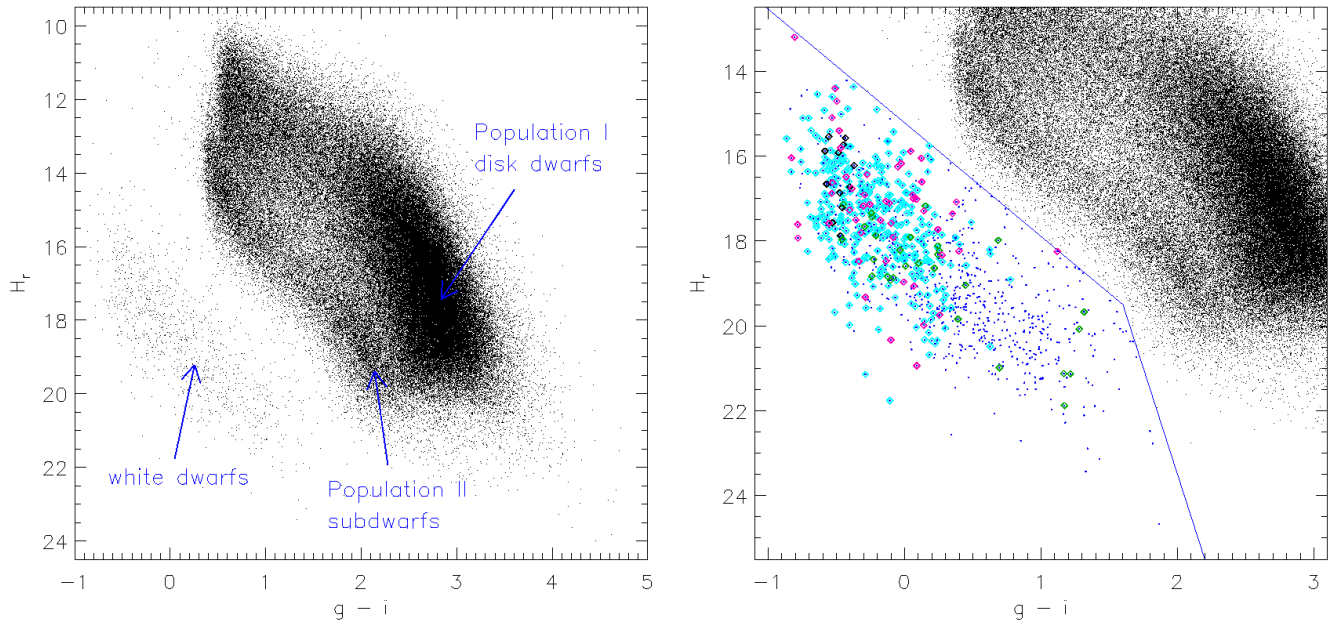}
\caption{Left: Reduced proper motion (RPM) diagram for Stripe 82. Plotted 
	magnitudes are inverse-variance weighted mean light curve magnitudes without 
	any correction for reddening. The sequences corresponding to white dwarfs, 
	Population I disk dwarfs, and Population II main sequence subdwarfs are
	well distinguishable. Right: Zoom on the white dwarf region of the 
	RPM diagram. The thin blue line separates the white dwarfs (blue dots on
	the left) from the dwarf and subdwarf populations (black dots on the right). 
	Previously spectroscopically confirmed white dwarfs are also shown: WDs 
	of spectral type DA (338 in total) as cyan diamonds, WDs of spectral type DB 
	(12 in total) as black diamonds, WDs of spectral type DC (29 in total) as 
	green diamonds and WDs of any other spectral type (50 in total) 
	as magenta diamonds~\citep{Ha03, Mc03, Kl04, Ca06, Ei06, Ki06, Si06}.
	\label{fig:rpm}}
\end{center}
\end{figure*}

SDSS Stripe 82 covers a $\sim 250$\,deg$^2$ area of sky, consisting of a
$2.5^\circ$  strip along the celestial equator from right ascension 
$-49.5^\circ$ to $+49.5^\circ$.
The stripe has been repeatedly imaged between June and December each year from 
1998 to 2005. Sixty-two of the total of 134 available imaging runs were obtained
in 2005. This multi-epoch 5-filter photometric data set has been utilised 
by~\citet{Br07} to construct the Light-Motion-Curve Catalogue (LMCC) and the 
Higher-Level Catalogue (HLC)\footnote{The LMCC and the HLC will be publicly 
released as soon as the~\citet{Br07} paper, currently in preparation, is 
published.}. The LMCC contains 3\,700\,548 light-motion curves,
extending to magnitude 21.5 in $u$, $g$, $r$, $i$ and to magnitude 20.5 in $z$. 
A typical light-motion curve consists of $\simeq$30 epochs over a baseline of 
6--7 years. The root-mean-square (RMS) scatter in the individual position 
measurements is $23\,\mathrm{mas}$ in each coordinate for $r \le 19.0$, 
increasing exponentially to $60\,\mathrm{mas}$ at $r=21.0$~\citep{Br07}. The HLC
includes 235 derived quantities, such as mean magnitudes, photometric 
variability parameters and proper-motion, for each light-motion curve in the 
LMCC. As an illustration of the catalogue's potential, Fig.~\ref{fig:pm} shows a
measured proper motion in RA and Dec for SDSS J224845.93--005407.0, one of the 
ultracool white dwarf candidates extracted from the LMCC. The source is faint 
($r = 20.56$) and has a measured proper-motion of 
$204 \pm 5\,\mathrm{mas}\,\mathrm{yr^{-1}}$.

UKIDSS is the UKIRT Infrared Deep Sky Survey~\citep{La07,He06,Ir07,Ha07}, 
carried out using the Wide Field Camera~\citep{Ca07} installed on the United 
Kingdom Infrared Telescope (UKIRT). The survey is now well underway with three 
ESO-wide releases: the Early Data Release (EDR) in February 2006~\citep{Dy06}, 
the Data Release 1 (DR1) in July 2006~\citep{Wa07a} and the Data Release 2 in 
March 2007~\citep{Wa07b}. In fact UKIDSS is made up of five surveys. The UKIDSS 
Large Area Survey (LAS) is a near-infrared counterpart of the SDSS photometric 
survey. The UKIDSS DR2~\citep{Wa07b} provides partial coverage, consisting of 
observations in at least one of $YJHK$-bands~\citep{He06}, of $\sim 2/3$ of 
Stripe 82 to a depth of $K\simeq 20.1$.

\section{Data Analysis}

\subsection{A Reduced Proper Motion Diagram}

The reduced proper motion (RPM) is defined as $H=m+5\log\mu +5$, where $m$ is 
the apparent magnitude and $\mu$ is the proper motion in arcseconds per year. 
Recently,~\citet{Ki06} have combined the SDSS Data Release 2 (DR2) and the 
USNO--B catalogues, using the resulting distribution of RPMs as a tracer of cool
white dwarfs. Here, our HLC for Stripe 82 enables us to construct an RPM diagram
using only SDSS data. Magnitudes used in the RPM diagram are mean magnitudes, 
calculated from all available single epoch SDSS measurements. These values
therefore differ from magnitudes quoted in the publicly available SDSS database
which are based on a single photometric measurement. Differences are usually
small, of the order of some hundredths of a magnitude. Typical proper motion 
errors in right ascension and declination are $\sim 2\,\mathrm{mas\,yr^{-1}}$ 
for $r\sim 17$ and $\sim 3\,\mathrm{mas\,yr^{-1}}$ for $r\sim 21$ which is 
slightly better than that of~\citet{Ki06} despite the much shorter 
time baseline used for calculating proper motions in the HLC. Additionally, the 
shorter time baseline with many position measurements has the advantage of 
reducing the contamination due to mismatches of objects with larger proper 
motions. The Stripe 82 photometric catalogue extends approximately $1.5$ mag 
deeper than the SDSS-DR2/USNO-B catalogue, corresponding to a factor of two in 
distance. Taking into account the different sky coverage of the two catalogues, 
the resultant volume over which white dwarfs may be detected in the HLC is 
$\simeq 60\,$per cent of that accessible in the SDSS-DR2/USNO-B catalogue.

In the left panel of Fig.~\ref{fig:rpm}, we present the RPM diagram for all 
131\,398 objects in Stripe 82 that meet the following criteria: i) the light 
curve consists of at least nine epochs in the $g$, $r$, $i$ and $z$ filters, 
ii) the object is classified as stellar by the SDSS photometric analysis in at 
least 80\,per cent of the epochs, iii) the proper motion is measured with a 
$\mathrm{\sqrt{\Delta \chi^2}}\geq 9$. In the case of high proper motion objects
($\mu \geq 50\ \mathrm{mas}\ \mathrm{yr^{-1}}$), the latter requirement is 
relaxed to allow a $\mathrm{\sqrt{\Delta \chi^2}}\geq 6$ measurement. The delta
chi-squared $\Delta \chi^2$ of the proper motion fit is defined as:
\begin{equation}
\Delta \chi^2 = \chi^2_{\alpha,con}-\chi^2_{\alpha,lin}+
\chi^2_{\delta,con}-\chi^2_{\delta,lin}\quad ,
\end{equation}
where $\chi^2_{con}$ is the chi-squared of the $\alpha/\delta$ measurements for
a model that includes only a mean position and  $\chi^2_{lin}$ is the 
chi-squared of the $\alpha/\delta$ measurements for a model that includes a 
mean position and a proper motion. This $\Delta \chi^2$ statistic follows a
chi-square distribution with two degrees of freedom. A relatively high 
$\mathrm{\Delta \chi^2}$ threshold was adopted in order to keep distinct stellar
populations in the RPM diagram cleanly separated.

The object sample was then matched with the UKIDSS DR2 catalogue using a search 
radius of 4\farcs0. Approximately 70\,per cent of the sample had at least one 
near-infrared detection. The unmatched fraction results from a combination of 
the incomplete coverage of Stripe 82 in the UKIDSS DR2 and a proportion of the 
faintest SDSS objects possessing infrared magnitudes below the UKIDSS LAS 
detection limits.

\begin{figure*}
\begin{center}
\includegraphics[width=1.0\hsize]{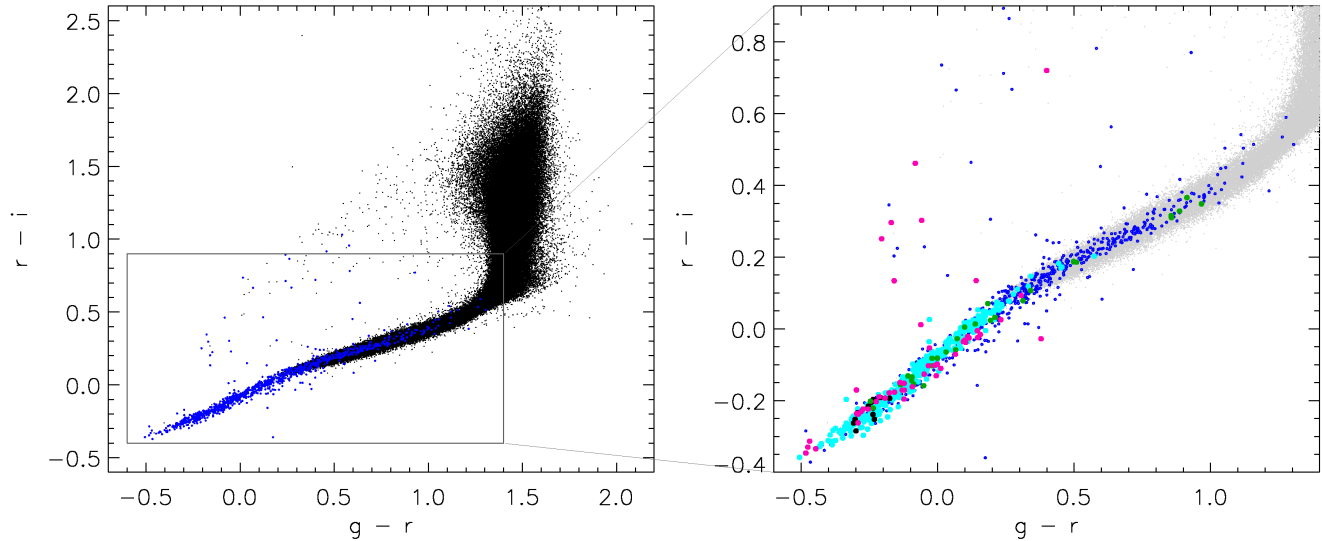}
\end{center}
\caption{Left: Colour-colour diagrams for all the objects plotted in the left 
	panel of Fig.~\ref{fig:rpm}. White dwarfs (candidates and
        confirmed) are overplotted as blue dots. The 
	overlap between the cooler end of the white dwarf locus and the main sequence
	star locus is clearly seen. Right: Zoom on the white dwarf region of the
	colour-colour diagram. The main sequence star locus is this
        time plotted in a 
	light grey colour. Spectroscopically confirmed white dwarfs
	are shown, using the same colour coding as in the right panel of
	Fig.~\ref{fig:rpm}. From this plot one can see that our new white dwarf 
	candidates are typically cooler objects.
\label{fig:cc}}
\end{figure*}

The left panel of Fig.~\ref{fig:rpm} shows three distinct sequences of stars,
namely Population I disk dwarfs, Population II main sequence subdwarfs and disk
white dwarfs. Given that our primary target population consists of nearby white
dwarfs, the magnitudes and colours used throughout the paper have not been
corrected for the effects of Galactic reddening. Adopting a boundary between
the subdwarfs and white dwarfs defined by $H_r > 2.68\,(g-i) + 15.21$ for 
$g-i \leq 1.6$ and $H_r > 10.0\,(g-i) + 3.5$ for $g-i >1.6$ produces a
sample of 1049 candidate white dwarfs (see the right panel of 
Fig.~\ref{fig:rpm}). 446 of the white dwarf candidates possess at least one 
detection in the near-infrared UKIDSS DR2. The discrimination boundary is 
similar to that employed by~\citet{Ki06} and lies in a sparsely populated 
region of the RPM diagram, where the individual magnitude and proper-motion 
errors are 
small\footnote{Typical error in $H_r$ is 0.25 and in $g-i$ is 0.02\,.}. As a 
consequence, the population of white dwarfs is not too sensitive to the precise
details of the separation boundary adopted. In this border region a small 
contamination from the subdwarfs is still possible~\citep{Ki06}.

In the left panel of Fig.~\ref{fig:cc} we show the overlap in colour 
space between the cooler end of the white dwarf locus with the main sequence 
star locus. One can see that it is impossible to distinguish cooler white 
dwarfs from the main sequence stars based on colour information only and proper 
motions for these cool objects are needed. The effectiveness of our RPM 
selection is evident from the location of 429 spectroscopically confirmed white
dwarfs from a number of sources~\citep{Ha03, Mc03, Kl04, Ca06, Ei06, Ki06, Si06}
in the right panels of Fig.~\ref{fig:rpm} and Fig.~\ref{fig:cc}. The previously 
confirmed white dwarfs populate practically only the upper half of the white 
dwarf sequence in the RPM diagram and the hotter end of the white dwarf locus 
in the colour-colour diagram. Confirmed cooler white dwarfs are rare and almost
all originate from~\citet{Ki06} where a similar detection method was adopted. 
The HLC is the faintest existing photometric/astrometric catalogue which 
enables us to trace hotter and thus intrinsically brighter white dwarfs to 
further distances and also to significantly enlarge the number of known cooler 
and thus intrinsically fainter white dwarfs.

Fig.~\ref{fig:rpm_candidates} is a copy of the right panel of Fig.~\ref{fig:rpm}
with the model loci for pure H atmosphere white dwarfs overplotted~\citep{Be01}.
For clarity only loci for $\log g=8.0$ and $\log g=9.0$, using typical disk and 
halo tangential velocities $\VT =30\,\mathrm{km s}^{-1}$ and 
$\VT=160\,\mathrm{km s}^{-1}$ are shown. In reality the $\log g$ values for 
white dwarfs range from $\sim 7.0$ to $\sim 9.5$ and the tangential velocities 
in the disk span from $\sim 20\,\mathrm{km s}^{-1}$ to 
$\sim 40\,\mathrm{km s}^{-1}$ which results in a strong overlap between 
different model loci. Because of this degeneracy it is impossible to determine 
unique white dwarf physical properties only from their positions in the RPM 
diagram. In the colour-colour diagrams of Fig.~\ref{fig:cc_candidates} the model
loci for pure H and He atmosphere white dwarfs are plotted for a range of log g 
values~\citep{Be01}. White dwarfs for different $\log g$ values have very 
similar colours. Colour differences are well measurable only at the very 
cool end of the model tracks. This holds for white dwarfs with either H or He
atmospheres. It is also not obviously feasible to distinguish between the H and He 
atmosphere white dwarfs. In some parts of the parameter space there is complete
degeneracy.

\subsection{$\chi^2$ Analysis}

In order to compare the available magnitude information for each white dwarf
candidate with the best matching model~\citep{Be01}, a minimal normalised 
$\chi^2$ value was calculated:
\begin{equation}
\chi^2=\frac{1}{N}\sum_{i=1}^{N}\left(\frac{m_i-m(T_{eff},\log{g})_{i}-D}{m_{err,i}}\right)^2\quad ,
\end{equation}
where $m_i$ are measured magnitudes for a given object, $m(T_{eff},\log{g})_{i}$
its corresponding model predictions and $m_{err,i}$ measured magnitude
errors\footnote{Magnitude errors smaller than 0.01 were replaced with this
minimal value in the $\chi^2$ calculation.}. The best solution was
sought on a grid of three free parameters, namely effective surface temperature 
($T_{eff}$), surface gravity ($\log{g}$) and distance modulus ($D$).
For each of the 1049 white dwarf candidates the four SDSS magnitudes, $g$, $r$, 
$i$ and $z$, were used in the fit. Whenever any of the UKIDSS magnitudes were 
available the fit was repeated with this additional information included. Only a
small subset of 70 white dwarf candidates have as yet all three $J$, $H$ and $K$ UKIDSS
magnitudes measured\footnote{$Y$ UKIDSS magnitude was not included in the
$\chi^2$ analysis due to the lack of white dwarf models for this 
waveband.}. Both H and He atmosphere models were compared with the data. In the 
attempt to better understand the degeneracy effects the best and also the 
second best solutions were recorded.

The statistical analysis of the obtained normalised $\chi^2$ results was 
performed first on the complete white dwarf sample, using only SDSS magnitude 
information. In all cases the $\chi^2$ distributions are strongly asymmetric. 
Typically fits of the H models perform better than the He ones. For the H 
models the normalised $\chi^2$ distribution peaks at $\sim 1.5$ and has a width 
of $\sim 3$ while for the He models the distribution peak is at $\sim 2.5$ and 
the width is $\sim 3.5$. Differences between the two fits are however usually 
too small to clearly distinguish between the two, especially in the temperature 
range where the H and He models are degenerate.

White dwarf atmosphere models observed in broad-band colours are also degenerate
for different surface gravity values. In order to estimate how strongly this 
degeneracy affects the results of the fit we compared the calculated best
normalised $\chi^2$ solutions with the second best ones. From the differences in
the calculated $\chi^2$ values it is usually not possible to completely discard
the second best solution. For instance, in the H model case, the second best
normalised $\chi^2$ distribution peaks at $\sim 3$ with a distribution width of
$\sim 3$. Nevertheless, for $\sim 60\,$per cent of the white dwarf candidates 
both best and second best solution predict the same effective temperature. For 
the remaining white dwarfs both temperatures usually differ from each other only
by $\leq 5\,$per cent, which typically corresponds to two adjacent bins on the 
discrete grid of available models. Available photometric information is thus 
sufficient to reliably estimate the white dwarf effective temperature. Not 
surprisingly, the fit results of the surface gravity (and as a consequence also
distance) are much less certain. The typical $\log{g}$ difference between the 
best and second best solution lies in the range from 0.5 to 1.5.

We used the subsample of 70 white dwarfs with complete UKIDSS photometry
available to show how the UKIDSS data improve the constraints on the model 
fits. The normalised $\chi^2$ of the fit using the complete SDSS and UKIDSS 
magnitude information is typically smaller by 0.5 than the value from the fit 
using SDSS data only. In $85\,$per cent of the cases the temperature predictions
are the same, otherwise the difference is only one step on the model temperature
grid, which confirms the reliability of the temperature estimation.
Moreover, in $85\,$per cent of the cases also the surface gravity solutions are
the same and in the remaining $15\,$per cent of the cases the best SDSS+UKIDSS 
solution often corresponds to the second best SDSS solution. This might 
indicate that the UKIDSS data help in breaking the degeneracy in the $\log{g}$ 
parameter. When this is not the case, the normalised $\chi^2$ are always larger
($>10$). 70 white dwarfs is still too small a number to draw definite 
conclusions, however, it seems that the inclusion of the UKIDSS data 
significantly enhances the credibility of the photometric fits. This will be 
very important in the near future, when it will be possible to match large 
portions of the SDSS/USNO-B measurements with the rapidly increasing LAS 
UKIDSS database.

\section{Candidates}

\subsection{Ultracool White Dwarfs}

\begin{figure*}
\begin{center}
\includegraphics[width=1.0\hsize]{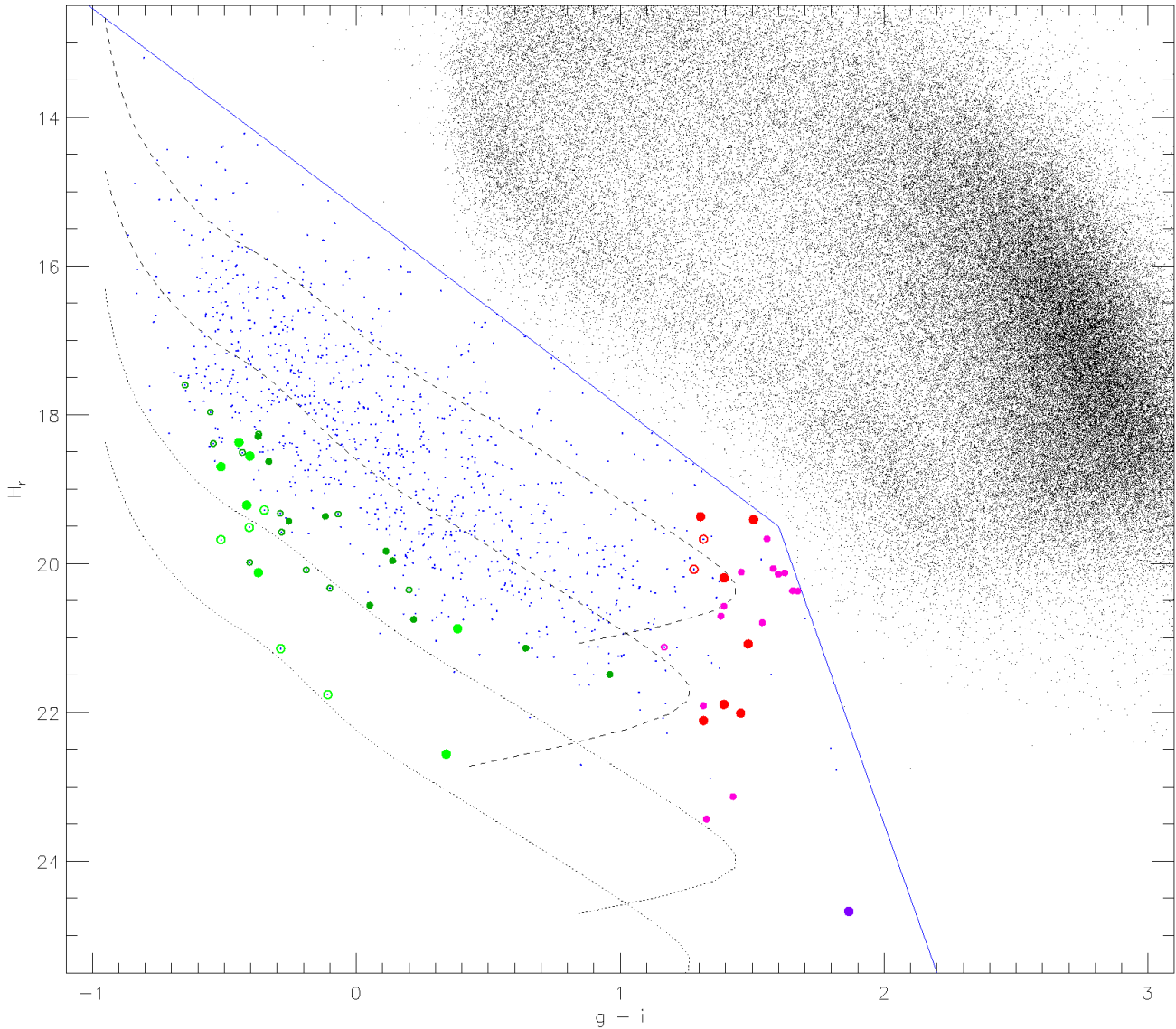}
\end{center}
\caption{As in the right panel of Fig.~\ref{fig:rpm} the zoom on the white dwarf
	region of the RPM diagram is shown. White dwarf H atmosphere
	models~\citep{Be01} for $\log g=8.0$ (upper of the dashed/dotted lines) 
	and $\log g=9.0$ (lower of the dashed/dotted lines), using typical disk
	and halo tangential velocities $\VT =30\,\mathrm{km s}^{-1}$ (dashed lines) 
	and $\VT=160\,\mathrm{km s}^{-1}$ (dotted lines) are overplotted.	Larger red 
	circles indicate H-rich ultracool white dwarf candidates, larger violet 
	circles He-rich ultracool white dwarf candidates and smaller magenta circles 
	possible H-rich ultracool white dwarf candidates (see detailed explanation in
	text). Halo white dwarf candidates are marked as light green larger circles 
	and possible halo white dwarf candidates as dark green smaller circles (see 
	detailed explanation in text). Empty circles of any colour mark objects that
	were already previously spectroscopically observed, full circles mark new 
	detections.
\label{fig:rpm_candidates}}
\end{figure*}

The signature of an ultracool surface temperature ($\Teff < 4000$\,K) is
depleted infrared flux, which makes the IR UKIDSS measurements crucial for the
recognition of the ultracool white dwarf candidates. Unfortunately only a 
small subsample of the 1049 white dwarfs have at least one UKIDSS measurement 
available as yet. That is why we first examined the outcome of the H and He
atmosphere model fits performed on the complete white dwarf sample, using SDSS 
magnitude information only. If the calculated surface 
temperature was smaller than $4000\,K$ and the normalised $\chi^2$ value of the 
fit was $< 15$ the object was added to the ultracool white dwarf candidate 
list. In the subsample of white dwarfs with at least one UKIDSS measurement this
additional piece of information was then used as a confirmation or rejection of 
potential ultracool white dwarf candidates. Typically the SDSS and UKIDSS 
predictions agreed.

The results of this analysis, graphically presented in
Fig.~\ref{fig:rpm_candidates} and Fig.~\ref{fig:cc_candidates}, are the 
following:
\begin{itemize}
\item[-] 9 ultracool white dwarf candidates with H-rich atmosphere among
which 7 have at least one supportive UKIDSS measurement. The object SDSS 
J224206.19$+$004822.7 has already been spectroscopically confirmed as a DC type
white dwarf (no strong spectral lines present, consistent with an H or He
atmosphere) by~\citet{Ki06} with a measured surface temperature of
$\sim 3400\,$K. The object SDSS J233055.19$+$002852.2 has been 
spectroscopically confirmed also as a DC type white dwarf by the same authors. 
Its measured surface temperature is $\sim 4100\,$K which is just above the
ultracool limit.
\item[-] One ultracool white dwarf candidate based on SDSS photometry only with 
He-rich atmosphere.
\item[-] 14 possible H-rich ultracool white dwarf candidates. For these
candidates the fit of the He atmosphere model gives in fact a slightly better 
result, predicting as a consequence an object with not so extremely low surface
temperature. Taking into account small differences between the H and He colours 
and the fact that H-rich white dwarfs are in fact much more numerous than the 
He-rich ones, makes these 14 objects still serious ultracool white dwarf 
candidates. The object SDSS J232115.67$+$010223.8 has been spectroscopically 
confirmed as a DC type white dwarf by~\citet{Ca06}. This 
object is also the only one where the fit based on only SDSS magnitudes did not 
predict an ultracool temperature but the inclusion of the UKIDSS K magnitude 
revealed the presence of the depleted IR flux. There are three additional 
ultracool candidates with at least one confirmed UKIDSS measurement.
\end{itemize}

Based on the positions in the RPM diagram and on the estimated tangential
velocities, we conclude that the vast majority of the newly discovered 
ultracool white dwarf candidates are likely members of the disk population. 
Measured and calculated properties of the 24 ultracool white dwarf candidates 
are presented in Table~\ref{tab:ultra}. Due to the relatively large temperature
errors, estimated to be at least $250\,K$ (the model temperature resolution in 
this temperature range), some of the 24 candidates might be in reality just 
above the ultracool white dwarf limit. We also do not quote the fit results for
the surface gravity since these are as explained above too uncertain. It is 
possible that some of the new ultracool white dwarf candidates close to the 
adopted boundary are subdwarfs instead. All these open issues can be ultimately
resolved only with spectroscopic follow up. However, already by analyzing the 
SDSS and UKIDSS astrometric/photometric information, it is certain that the 
total number of known ultracool white dwarfs has at least doubled.

\subsection{Halo White Dwarfs}

\begin{figure*}
\begin{center}
\includegraphics[width=1.0\hsize]{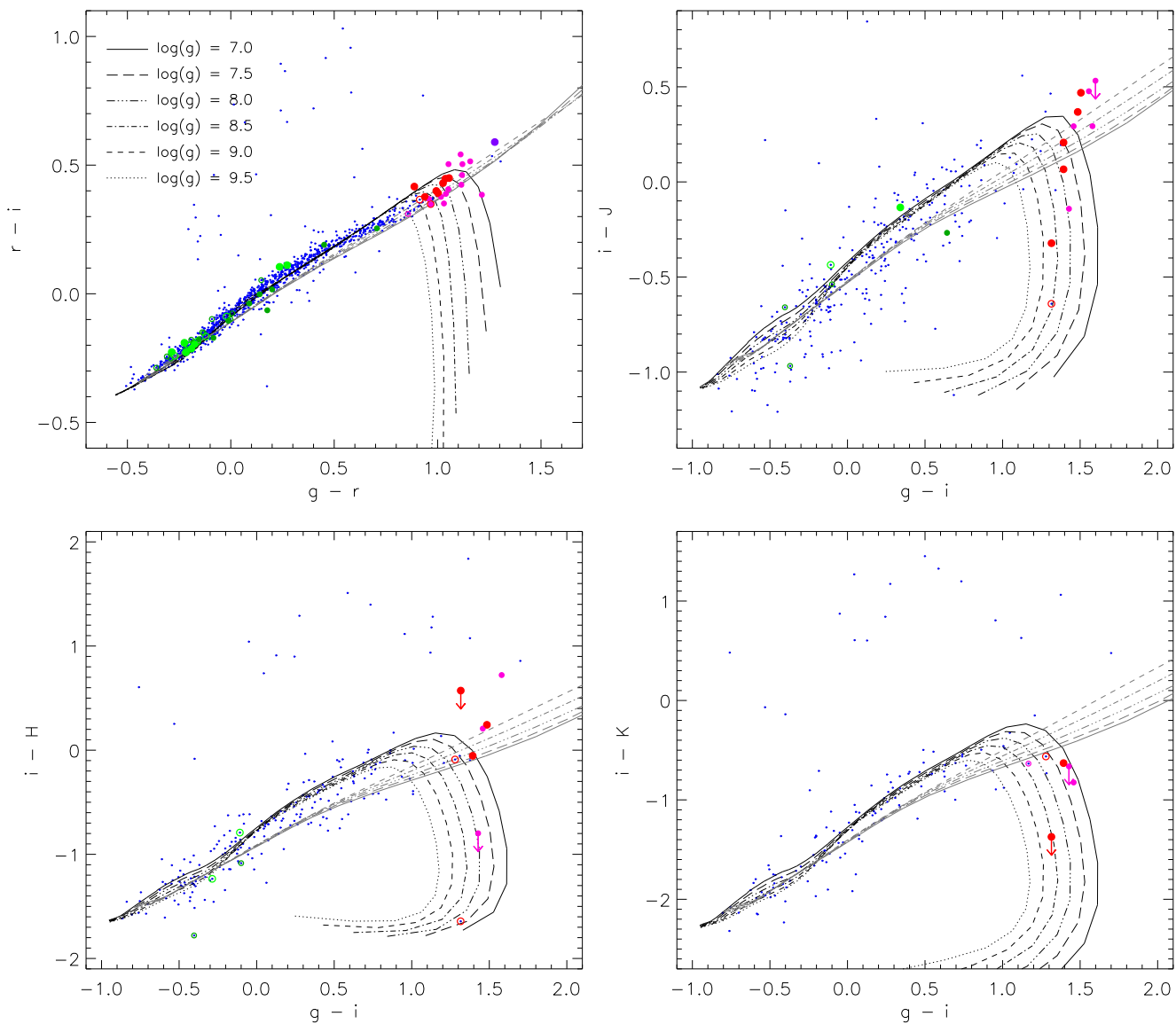}
\end{center}
\caption{Colour-colour diagrams for all the white dwarfs selected from the RPM
	diagram. H and He atmosphere white dwarf models~\citep{Be01} for different 
	surface gravity values are overplotted in black and grey respectively. Larger
	red circles indicate H-rich ultracool white dwarf candidates, larger violet 
	circle He-rich ultracool white dwarf candidate and smaller magenta circles 
	possible H-rich ultracool white dwarf candidates (see detailed explanation in
	text). The arrow indicates the limiting colour value for objects that were 
	not detected in J, H or K respectively. Halo white dwarf candidates are 
	marked as light green larger circles and possible halo white dwarf candidates
	as dark green smaller circles (see detailed explanation in text). Empty 
	circles of any colour mark objects that were already previously 
	spectroscopically observed, full circles mark new detections.
\label{fig:cc_candidates}}
\end{figure*}

\cite{Ha06} present a sample of 33 halo white dwarf candidates from their study
employing the SDSS DR3 and USNO-B catalogues. The combination of their larger
sky coverage and brighter magnitude limit of $r=19.7$ results in a survey volume
three times that of our Stripe 82 survey. 

Here we adopt the same criteria of $\VT >160\,\mathrm{km s}^{-1}$ as
\citet{Ha06} to select halo white dwarf candidates using the results of the
model fit. Unfortunately, white dwarf atmosphere models~\citep{Be01} are
degenerate for different values of the $\log g$ parameter in a
broad temperature range. This uncertainty in the calculation of the surface 
gravity is coupled with the determination of the distance to the object. 
Solutions with larger surface gravity and smaller distances are strongly coupled
with those with smaller surface gravity and greater distance. As a consequence 
determination of the tangential velocity is also somewhat uncertain.

Therefore we divide our halo white dwarf candidates, presented in
Table~\ref{tab:halo} and plotted in Fig.~\ref{fig:rpm_candidates} and 
Fig.~\ref{fig:cc_candidates}, into two groups:
\begin{itemize}
\item[-] white dwarfs where the best and also second best fit solution indicate
a quickly moving object. 12 highly probable halo candidates were thus found. 
Four were already previously spectroscopically observed~\citep{Kl04,Ei06} and 
two of them, namely SDSS J010207.24$-$003259.7 and SDSS J234110.13$+$003259.5 
were already included on the~\citet{Ha06} halo candidate list. 
\item[-] white dwarfs where the best fit solution gives 
$\VT >160\,\mathrm{km s}^{-1}$ but the second best fit solution does not. 22
white dwarfs were thus found and very likely not all of them are halo white
dwarfs. 12 were already previously spectroscopically 
observed by~\citep{Mc03,Kl04,Ei06}
\end{itemize}

One of the ultracool white dwarf candidates, SDSS J030144.09$-$004439.5 has its
calculated tangential velocity just below the threshold limit and could be thus
also a member of the halo population.

\section{Conclusions}

\citet{Br07} used the multi-epoch SDSS observations of Stripe 82 to build the 
deepest (complete to $r=21.5$) photometric/astrometric catalogue ever 
constructed, covering $\sim 250$\,deg$^2$ of the sky. Here, we have exploited 
the catalogue to find new candidate members of scarce white dwarf 
sub-populations. We extracted the stellar objects with high signal-to-noise 
ratio proper motions and built a RPM diagram. Cleanly separated from the 
subdwarf populations in the RPM diagram are 1049 cool white dwarf candidates.

Amongst these, we identified at least 7 new ultracool H-rich white dwarf 
candidates, in addition to one already spectroscopically confirmed 
by~\citet{Ki06} and one new ultracool He-rich white dwarf candidate . The 
effective temperatures $\Teff$ of these eight candidates are all $<$4000\,K. We
also identified at least 10 new halo white dwarf candidates with a tangential 
velocity of $\VT >160\,\mathrm{km s}^{-1}$. Spectroscopic follow-up of the new 
discoveries is underway.

In the quest for new members of still rare white dwarf sub-populations precise
kinematic information and even more the combination of SDSS and UKIDSS
photometric measurements will play an important role, as we illustrate
here. For the ultracool white dwarfs near IR UKIDSS measurements turn out to be
essential for making reliable identifications. In the near future the UKIDSS LAS
will cover large portions of the sky imaged previously by SDSS. This gives 
bright prospects for an even more extensive search for the still very scarce 
ultracool white dwarfs.

\section*{Acknowledgments}
We thank the referee, Nigel Hambly, for providing constructive comments and 
help in improving the contents of this paper. S. Vidrih acknowledge the 
financial support of the European Space Agency. L. Wyrzykowski was supported by
the European Community's Sixth Framework Marie Curie Research Training Network 
Programme, Contract No. MRTN-CT-2004-505183 "ANGLES". 
Funding for the SDSS and SDSS-II has been provided by the Alfred P. 
Sloan Foundation, the Participating Institutions, the National Science 
Foundation, the U.S. Department of Energy, the National Aeronautics and Space 
Administration, the Japanese Monbukagakusho, the Max Planck Society, and the
Higher Education Funding Council for England. The SDSS Web Site is 
http://www.sdss.org/.
                                                                               
The SDSS is managed by the Astrophysical Research Consortium for the 
Participating Institutions. The Participating Institutions are the American 
Museum of Natural History, Astrophysical Institute Potsdam, University of Basel,
Cambridge University, Case Western Reserve University, University of Chicago, 
Drexel University, Fermilab, the Institute for Advanced Study, the Japan 
Participation Group, Johns Hopkins University, the Joint Institute for Nuclear 
Astrophysics, the Kavli Institute for Particle Astrophysics and Cosmology, the 
Korean Scientist Group, the Chinese Academy of Sciences (LAMOST), Los Alamos 
National Laboratory, the Max-Planck-Institute for Astronomy (MPIA), the 
Max-Planck-Institute for Astrophysics (MPA), New Mexico State University, Ohio 
State University, University of Pittsburgh, University of Portsmouth, Princeton 
University, the United States Naval Observatory, and the University of 
Washington.

% Table 1
%\clearpage
\begin{table*}
\centering
\scriptsize
\rotcaption{Properties of the Ultracool White Dwarf Candidates}
\label{tab:ultra}
\begin{sideways}
\begin{tabular}[b]{l r@{$\,\pm\,$}l r@{$\,\pm\,$}l r@{$\,\pm\,$}l r@{$\,\pm\,$}l r@{$\,\pm\,$}l r@{$\,\pm\,$}l r@{$\,\pm\,$}l r@{$\,\pm\,$}l c c c c c c}
\hline
\multicolumn{1}{c}{Object} & \multicolumn{2}{c}{$g$} & \multicolumn{2}{c}{$r$} & \multicolumn{2}{c}{$i$} & \multicolumn{2}{c}{$z$} & \multicolumn{2}{c}{$J$} & \multicolumn{2}{c}{$H$} & \multicolumn{2}{c}{$K$} & \multicolumn{2}{c}{$\mu$} & $\varphi_\mu$ & $D$ & $V_T$ & $T_{eff}$ & $\chi^2$ & type \\
\multicolumn{1}{c}{SDSS J} & \multicolumn{2}{c}{\tiny[AB]} & \multicolumn{2}{c}{\tiny[AB]} & \multicolumn{2}{c}{\tiny[AB]} & \multicolumn{2}{c}{\tiny[AB]} & \multicolumn{2}{c}{\tiny[AB]$^{\dag}$} & \multicolumn{2}{c}{\tiny[AB]$^{\dag}$} & \multicolumn{2}{c}{\tiny[AB]$^{\dag}$} & \multicolumn{2}{c}{\tiny[mas yr$^{-1}$]} & \tiny[degree] & \tiny[pc] & \tiny[km s$^{-1}$] & \tiny[K] & \tiny{SDSS} & \\
\hline
001107.57$-$003102.8     & $21.68$ & $0.01$ & $20.62$ & $0.01$ & $20.17$ & $0.01$ & $19.87$ & $0.02$ & $19.71$ & $0.11$     & \multicolumn{2}{c}{}                 & \multicolumn{2}{c}{}                  & $ 57$ & $5 $ & $126$ & 125 &  34 & 3750 & 5.3 & H \\
004843.28$-$003820.0     & $22.55$ & $0.03$ & $21.44$ & $0.02$ & $20.97$ & $0.01$ & $20.53$ & $0.04$ & $20.68$ & $0.28$     & $20.25$ & $0.28$                     & \multicolumn{2}{c}{}		              & $ 53$ & $9 $ & $163$ & 165 &  42 & 3500 & 8.5 & H/He \\
012102.99$-$003833.6     & $20.74$ & $0.01$ & $19.75$ & $0.01$ & $19.35$ & $0.01$ & $19.17$ & $0.01$ & $19.14$ & $0.08$     & $19.40$ & $0.09$                     & $19.98$ & $0.19$                      & $123$ & $5 $ & $ 67$ &  55 &  32 & 3750 & 2.0 & H \\
013302.17$+$010201.3     & $22.74$ & $0.04$ & $21.62$ & $0.02$ & $21.11$ & $0.02$ & $20.56$ & $0.06$ & \multicolumn{2}{c}{} & \multicolumn{2}{c}{}                 & \multicolumn{2}{c}{}		              & $ 50$ & $9 $ & $272$ & 175 &  41 & 3500 &11.4 & H/He \\
030144.09$-$004439.5     & $20.45$ & $0.01$ & $19.41$ & $0.01$ & $19.02$ & $0.01$ & $18.81$ & $0.01$ & $19.16$ & $0.09$     & \multicolumn{2}{c}{$>19.81^{\ddag}$} & \multicolumn{2}{c}{$>19.68^{\ddag}$}  & $557$ & $3 $ & $169$ &  60 & 159 & 3750 & 9.0 & H/He \\
204332.97$+$011436.2     & $22.72$ & $0.04$ & $21.56$ & $0.02$ & $21.05$ & $0.02$ & $20.49$ & $0.07$ & \multicolumn{2}{c}{} & \multicolumn{2}{c}{}                 & \multicolumn{2}{c}{}	   	           & $ 58$ & $13$ & $122$ & 175 &  48 & 3500 & 8.9 & H/He \\
205010.17$+$003233.7     & $21.61$ & $0.04$ & $20.72$ & $0.06$ & $20.30$ & $0.10$ & $19.89$ & $0.06$ & \multicolumn{2}{c}{} & \multicolumn{2}{c}{}                 & \multicolumn{2}{c}{}                  & $ 54$ & $12$ & $193$ & 105 &  27 & 3750 & 1.4 & H \\
205132.05$+$000353.6     & $20.92$ & $0.02$ & $19.64$ & $0.02$ & $19.05$ & $0.02$ & $18.57$ & $0.03$ & \multicolumn{2}{c}{} & \multicolumn{2}{c}{}                 & \multicolumn{2}{c}{}		              &$1016$ & $13$ & $179$ &  20 &  88 & 3750 &13.1 & He \\
210742.26$-$002354.1     & $22.33$ & $0.03$ & $21.21$ & $0.01$ & $20.79$ & $0.01$ & $20.55$ & $0.05$ & \multicolumn{2}{c}{} & \multicolumn{2}{c}{}                 & \multicolumn{2}{c}{}	   	           & $ 82$ & $10$ & $132$ & 125 &  49 & 3500 & 1.9 & H/He \\
212216.01$-$010715.2     & $21.72$ & $0.02$ & $20.76$ & $0.01$ & $20.41$ & $0.01$ & $20.20$ & $0.04$ & \multicolumn{2}{c}{} & \multicolumn{2}{c}{}                 & \multicolumn{2}{c}{}                  & $170$ & $8 $ & $171$ &  65 &  51 & 3750 & 5.3 & H/He \\
212930.25$-$003411.5     & $21.24$ & $0.01$ & $20.28$ & $0.01$ & $19.91$ & $0.01$ & $19.72$ & $0.02$ & \multicolumn{2}{c}{} & \multicolumn{2}{c}{}                 & \multicolumn{2}{c}{}	   	           & $428$ & $5 $ & $180$ &  50 & 102 & 3750 & 8.5 & H/He \\
214108.42$+$002629.5     & $22.81$ & $0.04$ & $21.70$ & $0.02$ & $21.16$ & $0.02$ & $20.64$ & $0.06$ & \multicolumn{2}{c}{} & \multicolumn{2}{c}{}                 & \multicolumn{2}{c}{}	   	           & $ 54$ & $11$ & $178$ & 180 &  47 & 3500 &10.2 & H/He \\
220455.03$-$001750.6     & $21.58$ & $0.02$ & $20.55$ & $0.01$ & $20.20$ & $0.01$ & $19.98$ & $0.02$ & \multicolumn{2}{c}{} & \multicolumn{2}{c}{}                 & \multicolumn{2}{c}{}	   	           & $107$ & $6 $ & $188$ &  80 &  41 & 3750 & 7.1 & H/He \\
223105.29$+$004941.9     & $21.84$ & $0.01$ & $20.81$ & $0.01$ & $20.36$ & $0.01$ & $20.11$ & $0.03$ & $19.99$ & $0.12$     & $20.11$ & $0.17$                     & \multicolumn{2}{c}{}                  & $114$ & $5 $ & $233$ & 140 &  74 & 3750 & 7.3 & H \\
223520.19$-$003623.6     & $22.17$ & $0.02$ & $21.16$ & $0.01$ & $20.77$ & $0.01$ & $20.53$ & $0.04$ & $20.71$ & $0.28$     & \multicolumn{2}{c}{}                 & \multicolumn{2}{c}{}                  & $140$ & $6 $ & $ 81$ & 105 &  69 & 3750 & 1.6 & H \\
223715.32$-$002939.2     & $22.28$ & $0.03$ & $21.07$ & $0.01$ & $20.69$ & $0.01$ & $20.48$ & $0.04$ & \multicolumn{2}{c}{} & \multicolumn{2}{c}{$>20.15$}         & \multicolumn{2}{c}{}	   	           & $ 65$ & $7 $ & $122$ & 110 &  34 & 3250 & 9.6 & H/He \\
223954.07$+$001849.2     & $20.99$ & $0.05$ & $19.94$ & $0.02$ & $19.53$ & $0.01$ & $19.41$ & $0.02$ & $19.24$ & $0.06$     & $19.33$ & $0.10$                     & $20.36$ & $0.27$                      & $108$ & $4 $ & $352$ &  55 &  28 & 3500 & 0.5 & H/He \\
224206.19$+$004822.7$^a$ & $19.61$ & $0.01$ & $18.64$ & $0.01$ & $18.30$ & $0.01$ & $18.19$ & $0.01$ & $18.94$ & $0.05$     & $19.94$ & $0.15$                     & \multicolumn{2}{c}{$>19.70$}          & $161$ & $4 $ & $117$ &  20 &  17 & 3500 & 2.6 & H \\
224845.93$-$005407.0     & $21.50$ & $0.01$ & $20.56$ & $0.01$ & $20.19$ & $0.01$ & $19.98$ & $0.02$ & $20.51$ & $0.21$     & \multicolumn{2}{c}{$>19.64$}         & \multicolumn{2}{c}{}                  & $204$ & $5 $ & $112$ &  70 &  70 & 3750 & 1.3 & H \\
225244.51$+$000918.6     & $21.92$ & $0.02$ & $20.91$ & $0.01$ & $20.53$ & $0.01$ & $20.28$ & $0.05$ & \multicolumn{2}{c}{} & \multicolumn{2}{c}{}                 & \multicolumn{2}{c}{}	   	           & $ 86$ & $8 $ & $ 33$ & 120 &  49 & 3750 & 4.1 & H/He \\
232115.67$+$010223.8$^b$ & $19.81$ & $0.01$ & $18.96$ & $0.01$ & $18.65$ & $0.01$ & $18.49$ & $0.01$ & \multicolumn{2}{c}{} & \multicolumn{2}{c}{}                 & $19.28$ & $0.12$                      & $271$ & $6 $ & $201$ &  20 &  28 & 4250 & 8.3 & H/He \\
233055.19$+$002852.2$^c$ & $19.89$ & $0.01$ & $18.98$ & $0.01$ & $18.61$ & $0.01$ & $18.46$ & $0.01$ & \multicolumn{2}{c}{} & $18.70$ & $0.06$                     & $19.18$   & $0.09$                    & $166$ & $4 $ & $ 57$ &  25 &  21 & 3750 &11.2 & H \\
233818.56$-$004146.2     & $22.27$ & $0.02$ & $21.24$ & $0.01$ & $20.81$ & $0.01$ & $20.44$ & $0.04$ & \multicolumn{2}{c}{} & \multicolumn{2}{c}{}                 & \multicolumn{2}{c}{}		              & $142$ & $7 $ & $166$ & 165 & 112 & 3750 & 3.3 & H \\
234646.06$-$003527.6     & $22.52$ & $0.03$ & $21.46$ & $0.02$ & $20.96$ & $0.01$ & $20.53$ & $0.04$ & $20.48$ & $0.18$     & \multicolumn{2}{c}{}                 & \multicolumn{2}{c}{}	   	           & $ 44$ & $7 $ & $170$ & 165 &  34 & 3500 & 8.7 & H/He \\
\hline
\multicolumn{23}{r}{} \\
\multicolumn{23}{l}{\tiny{$^{a}$ DC type white dwarf with $T_{eff}\simeq 3400\,$K~\citep{Ki06}.}} \\
\multicolumn{23}{l}{\tiny{$^{b}$ DC type white dwarf~\citep{Ca06}.}} \\
\multicolumn{23}{l}{\tiny{$^{c}$ DC type white dwarf with $T_{eff}\simeq 4100\,$K~\citep{Ki06}.}} \\
\multicolumn{23}{r}{} \\
\multicolumn{23}{l}{\tiny{$^{\dag}$ The UKIDSS Vega magnitudes are converted to the AB system adopting a magnitude for Vega of +0.03 and the passband zero-point offsets from \citet{He06}.}}\\
\multicolumn{23}{l}{\tiny{$^{\ddag}$ 5$\sigma$ magnitude detection limit in J, H or K band respectively.}}
\end{tabular}
\end{sideways}
\end{table*}

 % Table 2
\clearpage
\begin{table*}
\centering
\scriptsize
\rotcaption{Properties of the Halo White Dwarf Candidates}
\label{tab:halo}
\begin{sideways}
\begin{tabular}[b]{l r@{$\,\pm\,$}l r@{$\,\pm\,$}l r@{$\,\pm\,$}l r@{$\,\pm\,$}l r@{$\,\pm\,$}l r@{$\,\pm\,$}l r@{$\,\pm\,$}l r@{$\,\pm\,$}l c c c c c c}
\hline
\multicolumn{1}{c}{Object} & \multicolumn{2}{c}{$g$} & \multicolumn{2}{c}{$r$} & \multicolumn{2}{c}{$i$} & \multicolumn{2}{c}{$z$} & \multicolumn{2}{c}{$J$} & \multicolumn{2}{c}{$H$} & \multicolumn{2}{c}{$K$} & \multicolumn{2}{c}{$\mu$} & $\varphi_\mu$ & $D$ & $V_T$ & $T_{eff}$ & $\chi^2$ & type \\
\multicolumn{1}{c}{SDSS J} & \multicolumn{2}{c}{\tiny[AB]} & \multicolumn{2}{c}{\tiny[AB]} & \multicolumn{2}{c}{\tiny[AB]} & \multicolumn{2}{c}{\tiny[AB]} & \multicolumn{2}{c}{\tiny[AB]$^{\dag}$} & \multicolumn{2}{c}{\tiny[AB]$^{\dag}$} & \multicolumn{2}{c}{\tiny[AB]$^{\dag}$} & \multicolumn{2}{c}{\tiny[mas yr$^{-1}$]} & \tiny[degree] & \tiny[pc] & \tiny[km s$^{-1}$] & \tiny[K] & \tiny{SDSS} & \\
\hline
000244.03$+$010945.8$^*$    & $19.96$ & $0.01$ & $20.11$ & $0.01$ & $20.29$ & $0.01$ & $20.35$ & $0.04$ & \multicolumn{2}{c}{} & \multicolumn{2}{c}{} & \multicolumn{2}{c}{} & $ 51$ & $5$ & $ 72$ & 760 & 182 & 10500 & 5.1 & H \\
000557.24$+$001833.2$^{*,a}$& $18.72$ & $0.01$ & $18.72$ & $0.01$ & $18.82$ & $0.01$ & $18.95$ & $0.01$ & $19.35$ & $0.10$ 	 & $19.90$ & $0.20$ 	   & \multicolumn{2}{c}{} & $210$ & $4$ & $ 92$ & 165 & 165 &  9000 & 0.6 & H \\
001306.23$+$005506.4$^{*,b}$& $19.34$ & $0.01$ & $19.47$ & $0.01$ & $19.63$ & $0.01$ & $19.76$ & $0.02$ & \multicolumn{2}{c}{} & \multicolumn{2}{c}{} & \multicolumn{2}{c}{} & $105$ & $4$ & $ 64$ & 525 & 261 & 10000 & 6.0 & H \\
001518.33$+$010549.2        & $19.80$ & $0.01$ & $20.03$ & $0.01$ & $20.22$ & $0.01$ & $20.39$ & $0.03$ & \multicolumn{2}{c}{} & \multicolumn{2}{c}{} & \multicolumn{2}{c}{} & $ 69$ & $5$ & $ 97$ & 795 & 259 & 11500 & 3.1 & H \\
001838.54$+$005943.5$^{*,c}$& $19.56$ & $0.01$ & $19.92$ & $0.01$ & $20.21$ & $0.01$ & $20.37$ & $0.04$ & \multicolumn{2}{c}{} & \multicolumn{2}{c}{} & \multicolumn{2}{c}{} & $ 34$ & $5$ & $156$ &1000 & 163 & 15500 & 4.3 & H \\
002951.64$+$005623.6	       & $20.25$ & $0.01$ & $20.43$ & $0.01$ & $20.62$ & $0.01$ & $20.67$ & $0.06$ & \multicolumn{2}{c}{} & \multicolumn{2}{c}{} & \multicolumn{2}{c}{} & $ 87$ & $6$ & $ 89$ & 525 & 216 & 12000 & 3.6 & H \\
003054.06$+$001115.6$^*$    & $20.75$ & $0.01$ & $20.66$ & $0.01$ & $20.69$ & $0.01$ & $20.49$ & $0.06$ & \multicolumn{2}{c}{} & \multicolumn{2}{c}{} & \multicolumn{2}{c}{} & $ 96$ & $8$ & $102$ & 455 & 207 &  8000 & 6.8 & H \\
003730.58$-$001657.8$^{*,d}$& $19.96$ & $0.01$ & $19.95$ & $0.01$ & $20.03$ & $0.01$ & $20.14$ & $0.03$ & \multicolumn{2}{c}{} & \multicolumn{2}{c}{} & \multicolumn{2}{c}{} & $ 75$ & $5$ & $217$ & 500 & 179 &  8500 & 1.3 & H \\
003813.53$-$000128.8$^{*,e}$& $18.91$ & $0.01$ & $19.10$ & $0.01$ & $19.31$ & $0.01$ & $19.55$ & $0.01$ & $19.97$ & $0.11$ 	 & $21.09$ & $ 0.35$	   & \multicolumn{2}{c}{} & $150$ & $4$ & $149$ & 500 & 357 & 11000 & 1.2 & H \\
004214.88$+$001135.7	       & $19.92$ & $0.01$ & $20.11$ & $0.01$ & $20.33$ & $0.01$ & $20.39$ & $0.05$ & \multicolumn{2}{c}{} & \multicolumn{2}{c}{} & \multicolumn{2}{c}{} & $ 49$ & $5$ & $ 92$ & 830 & 192 & 11500 & 4.2 & H \\
005906.78$+$001725.2$^{*,f}$& $19.13$ & $0.01$ & $19.27$ & $0.01$ & $19.42$ & $0.01$ & $19.60$ & $0.02$ & \multicolumn{2}{c}{} & \multicolumn{2}{c}{} & \multicolumn{2}{c}{} & $103$ & $4$ & $ 95$ & 480 & 233 & 10000 & 3.2 & H \\
010129.81$-$003041.7	       & $20.12$ & $0.01$ & $20.33$ & $0.01$ & $20.56$ & $0.01$ & $20.60$ & $0.05$ & \multicolumn{2}{c}{} & \multicolumn{2}{c}{} & \multicolumn{2}{c}{} & $ 40$ & $5$ & $118$ & 955 & 183 & 12000 & 5.4 & H \\
010207.24$-$003259.7$^g$    & $18.18$ & $0.01$ & $18.30$ & $0.01$ & $18.47$ & $0.01$ & $18.69$ & $0.01$ & \multicolumn{2}{c}{} & $19.70$ & $0.17$     & \multicolumn{2}{c}{} & $370$ & $3$ & $110$ & 125 & 221 & 11000 & 0.5 & H \\
010225.13$-$005458.4$^h$    & $19.28$ & $0.01$ & $19.55$ & $0.01$ & $19.79$ & $0.01$ & $20.06$ & $0.02$ & \multicolumn{2}{c}{} & \multicolumn{2}{c}{} & \multicolumn{2}{c}{} & $106$ & $4$ & $183$ & 525 & 264 & 13500 & 1.1 & H \\
014247.10$+$005228.4$^{*,i}$& $19.42$ & $0.01$ & $19.64$ & $0.01$ & $19.85$ & $0.01$ & $20.08$ & $0.02$ & \multicolumn{2}{c}{} & \multicolumn{2}{c}{} & \multicolumn{2}{c}{} & $ 59$ & $4$ & $ 51$ & 690 & 195 & 12000 & 1.7 & H \\
015227.57$-$002421.1$^*$    & $20.10$ & $0.01$ & $20.11$ & $0.01$ & $20.22$ & $0.01$ & $20.34$ & $0.03$ & \multicolumn{2}{c}{} & \multicolumn{2}{c}{} & \multicolumn{2}{c}{} & $ 71$ & $5$ & $111$ & 605 & 202 &  9000 & 0.5 & H \\
020241.81$-$005743.0$^*$    & $19.97$ & $0.01$ & $19.83$ & $0.01$ & $19.83$ & $0.01$ & $19.95$ & $0.02$ & \multicolumn{2}{c}{} & \multicolumn{2}{c}{} & $20.36$ & $0.27$	  & $106$ & $4$ & $124$ & 380 & 191 &  7500 & 1.3 & H \\
020729.85$+$000637.6$^*$    & $20.72$ & $0.01$ & $20.27$ & $0.01$ & $20.08$ & $0.01$ & $19.99$ & $0.02$ & $20.35$ & $0.14$ 	 & \multicolumn{2}{c}{} & \multicolumn{2}{c}{} & $149$ & $5$ & $ 88$ & 250 & 178 &  5500 & 4.6 & H \\
024529.69$-$004229.8	       & $19.88$ & $0.01$ & $20.16$ & $0.01$ & $20.39$ & $0.01$ & $20.56$ & $0.04$ & \multicolumn{2}{c}{} & \multicolumn{2}{c}{} & \multicolumn{2}{c}{} & $ 51$ & $4$ & $119$ & 690 & 167 & 13500 & 3.6 & H \\
024837.53$-$003123.9$^j$    & $19.22$ & $0.01$ & $19.24$ & $0.01$ & $19.32$ & $0.01$ & $19.41$ & $0.01$ & $19.76$ & $0.12$ 	 & $20.12$ & $0.19$ 	   & \multicolumn{2}{c}{} & $320$ & $4$ & $143$ & 145 & 219 &  9000 & 3.5 & H \\
025325.83$-$002751.5$^{*,k}$& $18.10$ & $0.01$ & $18.39$ & $0.01$ & $18.64$ & $0.01$ & $18.90$ & $0.01$ & \multicolumn{2}{c}{} & \multicolumn{2}{c}{} & $19.28$ & $0.12$	  & $100$ & $3$ & $ 82$ & 435 & 207 & 13500 & 0.3 & H \\
025531.00$-$005552.8$^l$    & $19.89$ & $0.01$ & $20.05$ & $0.01$ & $20.23$ & $0.01$ & $20.39$ & $0.03$ & \multicolumn{2}{c}{} & \multicolumn{2}{c}{} & \multicolumn{2}{c}{} & $ 70$ & $4$ & $ 88$ & 550 & 183 & 11000 & 5.7 & H \\
030433.61$-$002733.2	       & $21.20$ & $0.01$ & $20.93$ & $0.01$ & $20.82$ & $0.01$ & $20.67$ & $0.05$ & \multicolumn{2}{c}{} & $18.21$ & $0.03$     & $18.42$ & $0.04$  	  & $ 97$ & $7$ & $172$ & 480 & 221 &  6500 & 5.6 & H \\
211928.44$-$002632.9$^{*,m}$& $19.28$ & $0.01$ & $19.59$ & $0.01$ & $19.83$ & $0.01$ & $20.05$ & $0.03$ & \multicolumn{2}{c}{} & \multicolumn{2}{c}{} & \multicolumn{2}{c}{} & $ 47$ & $5$ & $ 73$ & 760 & 170 & 13500 & 2.7 & H \\
213641.39$+$010504.9$^*$    & $19.83$ & $0.01$ & $20.01$ & $0.01$ & $20.20$ & $0.01$ & $20.29$ & $0.04$ & \multicolumn{2}{c}{} & \multicolumn{2}{c}{} & \multicolumn{2}{c}{} & $ 45$ & $6$ & $190$ & 760 & 163 & 11000 & 4.5 & H \\
215138.09$+$003222.3$^{*,n}$& $20.38$ & $0.01$ & $20.23$ & $0.01$ & $20.18$ & $0.01$ & $20.20$ & $0.03$ & \multicolumn{2}{c}{} & \multicolumn{2}{c}{} & \multicolumn{2}{c}{} & $106$ & $6$ & $ 65$ & 400 & 199 &  7000 & 4.3 & H \\
223808.18$+$003247.9        & $20.51$ & $0.01$ & $20.28$ & $0.01$ & $20.17$ & $0.01$ & $20.08$ & $0.02$ & $20.30$ & $0.18$ 	 & \multicolumn{2}{c}{} & \multicolumn{2}{c}{} & $286$ & $6$ & $174$ & 140 & 187 &  6500 & 8.9 & H \\
223815.97$-$011336.9$^*$    & $20.76$ & $0.01$ & $20.59$ & $0.01$ & $20.65$ & $0.01$ & $20.61$ & $0.05$ & \multicolumn{2}{c}{} & \multicolumn{2}{c}{} & \multicolumn{2}{c}{} & $ 71$ & $7$ & $219$ & 550 & 184 &  7500 & 5.7 & H \\
230534.79$-$010225.2$^*$    & $20.08$ & $0.01$ & $20.17$ & $0.01$ & $20.34$ & $0.01$ & $20.42$ & $0.04$ & \multicolumn{2}{c}{} & \multicolumn{2}{c}{} & \multicolumn{2}{c}{} & $ 71$ & $5$ & $124$ & 525 & 177 & 10000 & 3.3 & H \\
231626.98$+$004607.0$^*$    & $21.15$ & $0.01$ & $20.44$ & $0.01$ & $20.19$ & $0.01$ & $20.07$ & $0.02$ & \multicolumn{2}{c}{} & \multicolumn{2}{c}{} & \multicolumn{2}{c}{} & $162$ & $5$ & $112$ & 220 & 168 &  5000 & 6.2 & H \\
233227.63$-$010713.8$^{*,o}$& $19.42$ & $0.01$ & $19.61$ & $0.01$ & $19.79$ & $0.01$ & $20.02$ & $0.02$ & $20.75$ & $0.27$ 	 & \multicolumn{2}{c}{} & \multicolumn{2}{c}{} & $ 54$ & $4$ & $236$ & 630 & 161 & 11000 & 1.1 & H \\
233817.06$-$005720.1$^{*,p}$& $20.34$ & $0.01$ & $20.43$ & $0.01$ & $20.53$ & $0.01$ & $20.58$ & $0.04$ & \multicolumn{2}{c}{} & \multicolumn{2}{c}{} & \multicolumn{2}{c}{} & $ 85$ & $6$ & $175$ & 760 & 307 &  9500 & 4.8 & H \\
234110.13$+$003259.5$^r$    & $19.12$ & $0.01$ & $19.32$ & $0.01$ & $19.52$ & $0.01$ & $19.73$ & $0.02$ & \multicolumn{2}{c}{} & \multicolumn{2}{c}{} & \multicolumn{2}{c}{} & $109$ & $4$ & $175$ & 575 & 298 & 11500 & 2.6 & H \\
235138.85$+$002716.9$^*$    & $21.04$ & $0.01$ & $20.84$ & $0.01$ & $20.82$ & $0.01$ & $20.70$ & $0.05$ & \multicolumn{2}{c}{} & \multicolumn{2}{c}{} & \multicolumn{2}{c}{} & $ 96$ & $8$ & $ 77$ & 400 & 181 &  7000 & 3.4 & H \\
\hline
\multicolumn{9}{l}{\tiny{$^{a}$ DZ type white dwarf~\citep{Kl04}.}}                                                      & \multicolumn{14}{l}{\tiny{$^{j}$ DA type white dwarf with $T_{eff}\simeq 9100\,$K and $log(g)\simeq 7.8$~\citep{Ei06}.}} \\
\multicolumn{9}{l}{\tiny{$^{b}$ DA type white dwarf with $T_{eff}\simeq 10900\,$K and $log(g)\simeq 8.1$~\citep{Ei06}.}} & \multicolumn{14}{l}{\tiny{$^{k}$ DA type white dwarf~\citep{Mc03}.}} \\
\multicolumn{9}{l}{\tiny{$^{c}$ DA type white dwarf with $T_{eff}\simeq 21800\,$K and $log(g)\simeq 7.6$~\citep{Ei06}.}} & \multicolumn{14}{l}{\tiny{$^{l}$ DA type white dwarf with $T_{eff}\simeq 12600\,$K and $log(g)\simeq 7.8$~\citep{Kl04}.}} \\
\multicolumn{9}{l}{\tiny{$^{d}$ DA type white dwarf with $T_{eff}\simeq 9100\,$K and $log(g)\simeq 7.9$~\citep{Ei06}.}}  & \multicolumn{14}{l}{\tiny{$^{m}$ DA type white dwarf with $T_{eff}\simeq 15700\,$K and $log(g)\simeq 7.9$~\citep{Ei06}.}} \\
\multicolumn{9}{l}{\tiny{$^{e}$ DA type white dwarf with $T_{eff}\simeq 12100\,$K and $log(g)\simeq 8.0$~\citep{Ei06}.}} & \multicolumn{14}{l}{\tiny{$^{n}$ DA type white dwarf with $T_{eff}\simeq 7800\,$K and $log(g)\simeq 8.2$~\citep{Ei06}.}} \\
\multicolumn{9}{l}{\tiny{$^{f}$ DZ type white dwarf~\citep{Kl04}.}}                                                      & \multicolumn{14}{l}{\tiny{$^{o}$ DA type white dwarf with $T_{eff}\simeq 14400\,$K and $log(g)\simeq 7.6$~\citep{Ei06}.}} \\
\multicolumn{9}{l}{\tiny{$^{g}$ DA type white dwarf with $T_{eff}\simeq 11100\,$K and $log(g)\simeq 8.2$~\citep{Kl04}.}} & \multicolumn{14}{l}{\tiny{$^{p}$ DA type white dwarf with $T_{eff}\simeq 10400\,$K and $log(g)\simeq 8.1$~\citep{Ei06}.}} \\
\multicolumn{9}{l}{\tiny{$^{h}$ DA type white dwarf with $T_{eff}\simeq 18800\,$K and $log(g)\simeq 9.0$~\citep{Ei06}.}} & \multicolumn{14}{l}{\tiny{$^{r}$ DA type white dwarf with $T_{eff}\simeq 13500\,$K and $log(g)\simeq 7.9$~\citep{Kl04}.}} \\
\multicolumn{9}{l}{\tiny{$^{i}$ DA type white dwarf with $T_{eff}\simeq 14000\,$K and $log(g)\simeq 8.8$~\citep{Ei06}.}} \\
\multicolumn{23}{l}{}\\
\multicolumn{23}{l}{\tiny{$^*$} Only the best and not also the second best solution predict a halo white dwarf candidate.}\\
\multicolumn{23}{l}{\tiny{$^{\dag}$ The UKIDSS Vega magnitudes are converted to the AB system adopting a magnitude for Vega of +0.03 and the passband zero-point offsets from \citet{He06}.}}
\end{tabular}
\end{sideways}
\end{table*}

\end{document}